\documentclass[showpacs,floatfix,preprint]{revtex4-1}
\usepackage{graphicx,psfrag,amsmath,amssymb,amsfonts,latexsym,color,epsf,graphpap,dcolumn}

\definecolor{red}{rgb}{1,0,0}
\definecolor{blue}{rgb}{0,0,1}
\definecolor{skyblue}{rgb}{0,0,.5}
\definecolor{green}{rgb}{0,1,0}
\definecolor{orange}{cmyk}{0,.4,1,0}

\newcommand{\fq}{\left(\frac{\hbar}{2 e}\right)^2}
\newcommand{\ELcav}{E_{L,\text{cav}}}

\begin{document}
\title{Dynamical Casimir effect in superconducting circuits:  a numerical approach }
\author{F.C.~Lombardo$^{a}$, F.D.~Mazzitelli$^{b}$, A. Soba $^c$, P.I.~Villar $^a$}
\affiliation{
$^a$Departamento de F\'\i sica {\it Juan Jos\'e
Giambiagi}, FCEyN UBA and IFIBA CONICET-UBA, Facultad de Ciencias Exactas y Naturales,
Ciudad Universitaria, Pabell\' on I, 1428 Buenos Aires, Argentina.\\
$^b$Centro At\'omico Bariloche and Instituto Balseiro, 
Comisi\'on Nacional de Energ\'\i a At\'omica, R8402AGP Bariloche, Argentina.\\
$^c$Centro de Simulaci\'on Computacional para Aplicaciones Tecnol\'ogicas, CSC - CONICET \\
Polo Cient\'\i fico y Tecnol\'ogico de Buenos Aires; Godoy Cruz 2390, Buenos Aires.}
\date{today}

\begin{abstract}
\noindent 
We present a numerical analysis of the particle creation for a quantum field in the presence of time dependent
boundary conditions. Having in mind recent experiments involving superconducting circuits, we consider their
description in terms of a scalar field in a one dimensional cavity  satisfying generalized boundary conditions that involve a time-dependent linear
combination of the field and its spatial and time derivatives. We evaluate numerically the Bogoliubov transformation
between {\it in} and {\it out}-states and find that the rate of particle production strongly depends on whether the  spectrum of
the unperturbed cavity is equidistant or not, and also on the amplitude of the temporal oscillations of the boundary conditions.
We provide analytic justifications for the different regimes found numerically.
\end{abstract}
\maketitle
\section{Introduction}\label{sec:intro}

Non-adiabatic time dependent external conditions can excite any quantum system. In the context of quantum field 
theory, the initial vacuum state generally evolves into an excited state with a non-vanishing number of particles \cite{Moore}. For
example,  time dependent gravitational or electromagnetic fields can induce particle creation. The same phenomena take place
in the presence of time-dependent environments, as for instance a  cavity with time dependent size or  electromagnetic properties.
The latter type of situations are broadly named `dynamical Casimir effect" (DCE). For recent reviews see \cite{Dodonov2010, Dalvit2011, 
Nation2012}.

A large body of the literature on this subject is devoted to the study of particle creation in the presence of `moving mirrors", which 
 impose boundary conditions at its position. When accelerated, the mirror excites the quantum state of the electromagnetic field,
and therefore create photons from an initial vacuum state. The rate of particle production is generally very small. However, it  can be notably enhanced in a closed cavity at the hand of parametric resonance, when the length of the cavity oscillates with a frequency $\Omega = 2 \omega_n$ where $\omega_n$ is one of  the eigenfrequencies of the static cavity \cite{param1,param2}. Even in this case, the total number of created photons is severely restricted 
by the $Q$-factor of the cavity, and the experimental verification of this phenomenon is still a challenge \cite{Dodonov2010, Dalvit2011}.

Partly due to these reasons, there have been alternative proposals to verify experimentally the particle creation in the presence of time 
dependent environments, as suggested earlier in Ref.\cite{Yablo}. 
For instance, a thin semiconductor sheet inserted in a closed cavity can induce particle creation when its conductivity
is rapidly changed with ultra short  laser pulses. The effective length of the cavity changes suddenly each time the conductivity changes, provided the time
dependent boundary condition is similar to that of a moving mirror. Although there are some experimental advances on this proposal, this effect could not be 
measured yet \cite{Braggio}.   Another interesting possibility is to consider time dependent refractive index perturbations produced by intense laser pulses 
in optical fibers or thin materials \cite{Belgiorno}, which can be used to construct models that mimic phenomena of semiclassical  gravity like Hawking radiation, Unruh
effect, or cosmological particle creation \cite{analogue}.

An alternative setup, closely related to the main topic of the present paper,  consists of  a superconducting waveguide ended with a 
superconducting quantum interference device (SQUID), which determines the 
boundary condition of the field at that point. 
A time dependent magnetic flux through the SQUID generates a time dependent boundary condition, with the subsequent excitation of the
field (particle creation) in the waveguide \cite{Nori}.   A few years ago, the DCE has been experimentally observed using this system  \cite{Wilson}. 
An array of SQUIDs can also be used to measure the DCE, simulating not a moving mirror but a time dependent refraction index, see \cite{Paraoanu}.

A natural generalization of the proposal of Ref.\cite{Nori} is to consider
a superconducting cavity of finite size (for example a waveguide ended with two SQUIDs). The electromagnetic field inside the cavity can be described by a single quantum 
massless scalar field $\phi(x,t)$, where $x$ is the spatial coordinate along the waveguide. The field satisfies the wave equation in the cavity, along with
the boundary conditions imposed by the SQUIDs.  Assuming that the SQUIDs are
located at $x=0$ and $x=d$, and that a time dependent magnetic flux is applied only at $x=d$, the boundary conditions are of the form
\begin{eqnarray}\label{bc1}
&&\phi ' (0,t) + A_1 \phi(0,t)+ B_1 \ddot\phi(0,t) = 0\nonumber\\
&&\phi ' (d,t) + A_2(t) \phi(d,t)+ B_2(t)\ddot\phi(d,t) = 0\, 
\end{eqnarray}
where primes and dots denote derivatives with respect to $x$ and $t$, respectively. The constants $A_1$ and $B_1$ are determined by the physical
properties of the cavity while the functions $A_2(t)$ and $B_2(t)$ also depend on the time dependent external magnetic flux.  
In a `static" situation (when $A_2$ and $B_2$ are constants), the eigenfrequencies form a discrete set, and therefore particle creation can be 
increased through parametric resonance, by choosing the frequency of the external perturbation.

From a mathematical 
point of view, the system is therefore modeled by a massless scalar field in $1+1$ dimensions satisfying generalized Robin boundary conditions. The generalization 
involves not only time dependent parameters, but also the presence of the second time derivative $\ddot \phi$. This term indicates that, in fact, there is a 
degree of freedom localized at $x=d$ (the mechanical analog would be a string ended with a mass, the boundary condition being the equation of motion of the mass,
see for instance \cite{Fosco2013}). If the time dependence of the boundary conditions starts at $t=0$, the constant values of $A_i$ and $B_i$ for $t<0$
determine the spectrum of the static cavity, that can be tuned setting different values for the properties of the SQUIDs.
 
We are thus led to the problem of analyzing particle creation for a scalar field in $1+1$ dimensions subjected to the boundary conditions Eq.(\ref{bc1}).  
This problem has been partially addressed in previous works.  Ref.\cite{Farinaetal} considered the case of a single mirror with time dependent
Robin boundary conditions, and computed perturbatively the spectrum of created 
particles. In Ref.\cite{Louko2015} a similar problem was studied analytically, considering non trivial boundary conditions at one or two points, and 
paying particular attention on  whether time dependent Robin parameter reproduces the boundary condition for a moving mirror or not. 
The effects produced by the inclusion of the term proportional to $\ddot\phi$, not considered in \cite{Louko2015},
have been analyzed for waveguides ended by a SQUID  in \cite{Andreson}.  Some of us \cite{Fosco2013} considered the case of a closed cavity with 
general boundary conditions, showing that parametric resonance may induce an exponential growth in the number of particles inside the cavity.   

In the present paper we will present a detailed numerical analysis of the particle creation rate, along with analytical calculations that describe the main features
of the numerical results (for previous numerical approaches in the case of Dirichlet or Neumann boundary conditions see \cite{numerical}).
As we will see, particle creation depends crucially on the properties of the static spectrum of the cavity, and also on the amplitude
of the time dependent part of the functions $A_2(t)$ and $B_2(t)$.   This is to be expected from previous results for the case of moving mirrors that impose Dirichlet  or 
Neumann boundary conditions. Indeed, when the spectrum is not equidistant, parametric resonance involves a finite number of modes, and the number of particles in these modes grow exponentially when the external frequency is properly chosen \cite{crocces}. On the other hand, for equidistant spectra, the coupling between an infinite number of modes makes the number of particles to grow quadratically or linearly in time, for short or long time-scales respectively \cite{param1}, while the total energy grows exponentially.  We will show that similar situations hold for the superconducting cavities. Moreover, the numerical approach will allow us to address cases where the boundary conditions oscillate with a large amplitude 
(that are in principle accessible for experiments)  but do not admit an analytical description.

The paper is organized as follows.
 In Section 2 we describe the model for a (linearized) superconducting cavity  with time dependent boundary conditions, and show that the system can be described as a set of coupled harmonic oscillators. In Section 3 we study analytically  the particle creation rate using multiple scale analysis (MSA), emphasizing the dependence of the results
 with the main characteristics of the spectrum. Section 4 contains the numerical results, along with a brief discussion of the numerical method.
We compute the particle creation rate for spectra of different characteristics, and compare numerical and analytical results. Section 5 contains a discussion of the results and the main conclusions of our work.


\section{Tunable superconducting cavity}


We shall consider a superconducting cavity of length $d$ which is decoupled from the input line  at $x=0$,  and has a SQUID at $x=d$. For the theoretical description we 
follow closely Ref.\cite{Shumeiko}.
The cavity, which is assumed to have capacitance $C_0$ and inductance $L_0$ per unit length,   is described by the superconducting phase field $\phi(x,t)$ with Lagrangian
\begin{equation}\label{lag}
L_{\rm cav} = \fq \frac{C_0}{2} \int_0^d d x \left(\dot \phi^2 - v^2 \phi'^2 \right) 
+ \left[ \fq \frac{2 C_J}{2} \dot \phi_d^2 
 -  E_J \cos{f(t)} \phi_d^2
 \right]
 \,,
\end{equation}
where  $v = 1/\sqrt{L_0 C_0}$ is the field propagation velocity,  $\phi_d$ the value of the field at the boundary by $\phi(d,t)$, and $f(t)$ 
is the phase across the SQUID controlled by external magnetic flux. $E_J$ and $C_J$ denote the Josephson energy and capacitance, respectively. 
The Lagrangian in Eq.(\ref{lag}) contains additional contributions proportional to higher powers of $\phi_d$ that  will not be considered in the rest of this paper.

As anticipated, the description of the cavity involves the field $\phi(x,t)$ for $0<x<d$ and the  additional degree of freedom $\phi_d$. The dynamical equations read
\begin{equation}
\ddot \phi - v^2 \phi'' = 0
\,,
\end{equation}
and
\begin{equation}\label{eqphid}
\frac{\hbar^2}{E_C} \ddot \phi_d + 2 E_J \cos{f(t)} \phi_d + \ELcav d \phi'_d  = 0
\,,
\end{equation}
where $E_C = (2e)^2/(2 C_J)$ and $\ELcav = (\hbar/2e)^2 (1/L_0 d)$. The equation above comes from the variation of the action with respect to $\phi_d$,
and can be considered as a generalized boundary condition for the field.   We could consider general boundary conditions also at $x=0$, but for the sake of simplicity  we will assume
that  $\phi'(0,t)=0$ (physically this corresponds to the situation where the cavity is decoupled).

It will be useful to write the Lagrangian in terms of eigenfunctions of the static cavity. Assuming that
\begin{equation}
f(t)=f_0 +  \theta (t) \theta(t_F-t)\epsilon\sin\Omega t\, ,\label{pert}
\end{equation}
we can expand the field as
\begin{equation}\label{modeexpansion}
\phi(x,t) = {2e \over \hbar} \sqrt{2 \over C_0 d} \sum_n q_n(t) \cos k_nx
\,,
\end{equation}
where the eigenfrequencies $k_n$ satisfy Eq.(\ref{eqphid}) in the static case $f=f_0$:
\begin{equation}
\label{spectrum}
(k_n d)\tan{k_n d}  =   \frac{2 E_J \cos{f_0} }{\ELcav}  - \frac{2 C_J }{C_0 d}(k_n d)^2,
\end{equation}
In terms of the new variables $q_n(t)$ the Lagrangian reads

\begin{equation}\label{lagqn}
 L_{\rm cav} 
= \frac{1}{2} \sum_n \left[ M_n \dot q_n^2 - M_n \omega_n^2 q_n^2 \right] 
+ E_J \phi_d^2 [\cos f_0-\cos f(t)] 
\, ,
\end{equation}
where $\omega_n = v k_n$ and 
\begin{equation}\label{Mn}
M_n = 1 + {\sin{2k_n d} \over 2k_nd} + {4C_J\over dC_0}\cos^2k_nd
\,.
\end{equation}
The dynamical  equation for the mode $n$ is therefore
\begin{equation}\label{eqmodes}
\ddot{q}_n + v^2 k_n^2 q_n = 
\frac{4 E_J}{\ELcav M_n}\frac{v^2}{d^2} \epsilon \, \theta(t)\theta(t_F-t) \sin (f_0) \sin\Omega t \cos k_n d \sum_m q_m(t) \cos k_md \, ,
\end{equation}
where we assumed that $\epsilon\ll 1$.

The classical description of the theory consists of a set of coupled harmonic oscillators
with time dependent frequencies. The quantization of the system is straightforward. In the Heisenberg representation,
the variables $q_n(t)$ become quantum operators
\begin{equation}
\hat q_n(t)=u_n(t)\hat a_n + u_n^*(t)\hat a^{\dagger}_n\, , \label{expansion}
\end{equation}
where $\hat a_n$ and  $\hat a^{\dagger}_n$ are the annihilation and creation operators. The functions $u_n(t)$ are properly normalized solutions 
of Eq.(\ref{eqmodes}). In the static regions $t<0$ and $t>t_F$ they are linear combinations of $e^{\pm i v k_n t}$.  We define the {\it in}-basis as the 
solutions of Eq.(\ref{eqmodes}) that satisfy
\begin{equation}
u_n^{in}(t)=\frac{1}{\sqrt{2 v k_n}}e^{-i v k_n t} \,\,\, \text {for}\,\,\,  t<0 \,\, .
\end{equation}
The associated annihilation operators $a_n^{in}$ define the {\it in}-vacuum $\vert 0_{in}\rangle$. The {\it out }-basis 
$u_n^{out}$ is introduced in a similar way, defining the behavior  for $t>t_F$. 
The {\it in} and {\it out} basis are connected by a Bogoliubov transformation
\begin{equation}
u_n^{in}(t)=\alpha_n u_n^{out}(t)+\beta_n u_n^{out\, *}(t)\, ,
\end{equation}
and the number of created particles in the mode $n$ for $t>t_F$ is given by
\begin{equation}
N_n=\langle 0_{in}\vert a_n^{out\, \dagger} a_n^{out}\vert0_{in}\rangle =\vert\beta_n\vert^2\, . \label{Npart}
\end {equation}

In the present paper, we shall numerically solve the dynamical Eqs.(\ref{eqmodes}) and evaluate the number of created particles using (\ref{expansion}) to (\ref{Npart}).

\section{Some analytic results: multiple scale analysis}

In order to study analytically  Eqs.(\ref{eqmodes}) we write them in the form
\begin{equation}\label{eqmodecomp}
\ddot q_n +\omega_n^2(t)q_n=\sum_{m\neq n}S_{nm}(t)q_m\, ,
\end{equation}
where we made the redefinition $q_n\to \sqrt{M_n}q_n$ and
\begin{eqnarray}
\omega_n(t)&=&k_n(1-\alpha \frac{k_1^2}{k_n^2}\frac{\cos^2k_n d}{M_n} \sin\Omega t)\nonumber\\
S_{mn}(t)&=&  \alpha k_1^2 \frac{\cos k_n d\cos k_m d}{\sqrt{M_nM_m}} \sin\Omega t\nonumber\\
\alpha&=& \frac{4 E_J}{\ELcav k_1^2 d^2} \epsilon \, \sin (f_0) \, .
\label{ec16}
\end{eqnarray} 
Here $k_1$ denotes the lowest eigenfrequency \cite{aclar}. We will assume that the amplitude of the 
time dependence is small, that is $\alpha\ll 1$.  We have set $v = 1$ and shall use this in what follows.  

It is known that, due to parametric resonance,  a naive perturbative solution of Eqs.(\ref{eqmodecomp})
in powers of $\alpha$
breaks down after a short amount of time.
In order  to find a solution valid for longer times  we use the multiple scale analysis (MSA) technique \cite{bender, crocces}. 
We introduce a second timescale $\tau=\alpha t$, and write
\begin{equation}
q_n(t,\tau)=A_n(\tau)\frac{e^{i k_n t}}{\sqrt{2 k_n}}+B_n(\tau)\frac{e^{-i k_n t}}{\sqrt{2 k_n}}\, . \label{ec17}
\end{equation}
The functions $A_n$ and $B_n$ are slowly varying, and contain the cumulative resonant effects. To obtain differential equations for them, we insert 
this ansatz into Eq.(\ref{eqmodecomp}) and neglect second derivatives of $A_n$ and $B_n$. After multiplying the equation by 
$\exp{\pm i k_n t}$, and averaging over the fast oscillations we obtain
\begin{eqnarray}\label{eqsAyB}
\frac{dA_n}{d\tau}&=& - \frac{k_1^2}{k_n}\frac{\cos^2k_n d}{2 M_n}B_n\delta(\Omega-2k_n)+ \frac{k_1^2}{4}\sum_{m\neq n}\frac{\cos k_n d\cos k_md}{\sqrt{k_nk_mM_nM_m}}\nonumber\\
 &\times & \left [  A_m\left(\delta(k_n-k_m+\Omega)-\delta (k_n-k_m-\Omega)\right)-B_m\delta(k_n+k_m-\Omega)\right],\nonumber\\
\frac{dB_n}{d\tau}&=& - \frac{k_1^2}{k_n}\frac{\cos^2k_n d}{2 M_n}A_n\delta(\Omega-2k_n)- \frac{k_1^2}{4}\sum_{m\neq n}\frac{\cos k_n d\cos k_md}{\sqrt{k_nk_mM_nM_m}}\nonumber\\
 &\times & \left [  B_m\left(\delta(k_m-k_n+\Omega)-\delta (k_m-k_n-\Omega)\right) + A_m\delta(k_n+k_m-\Omega)\right]\, .
\end{eqnarray}
We can see that these equations are non trivial when the
 external harmonic driving frequency is just tuned with one eigenvalue of the static cavity $\Omega= 2k_n$. Moreover, other modes will be coupled
 and will resonate if the condition 
 \begin{equation}\label{rescondition}
 \Omega = \vert k_n\pm k_j\vert
 \end{equation}
 is satisfied.

\subsection{A single resonant mode}

If we assume that Eq.(\ref{rescondition}) is not satisfied by any $j\neq n$, the only resonant mode is the one tuned with the external frequency.
In this case, the long time solution to Eqs.(\ref{eqsAyB}) is trivial and given by
\begin{equation}
\vert A_n\vert^2\simeq\vert B_n\vert^2\simeq e^{\lambda_n t}\, ,\quad \lambda_n=\alpha  \frac{k_1^2}{k_n}\frac{\cos^2k_n d}{ 2 M_n}\, ,
\label{onemode}\end{equation}
that is, due to parametric resonance the number of created particles grows exponentially with the stopping time.

\subsection {Finite number of resonant modes} 

Let us now consider the case in which Eq.(\ref{rescondition}) is satisfied for a finite number of modes. In this case, Eqs.(\ref{eqsAyB})
become nontrivial for that modes, and one expects an enhancement in the corresponding particle creation rates. Although we could consider a 
general situation, for the sake of simplicity we will illustrate this case with a particular example.  We will assume that the system is excited with an
 frequency $\Omega = 2 k_j$, and that exists another eigenfrequency of the static cavity such that $k_l=3 k_j$. Thus,   Eqs.(\ref{eqsAyB})
 become
 \begin{eqnarray}
 \frac{d A_j}{d\tau} &=& - \Gamma_j B_j + \Gamma_{jl} A_l\nonumber\\
  \frac{d B_j}{d\tau} &=& - \Gamma_j A_j + \Gamma_{jl} B_l\nonumber\\
 \frac{d A_l}{d\tau} &=& - \Gamma_{jl} A_j\nonumber\\
   \frac{d B_l}{d\tau} &=& -  \Gamma_{jl} B_i\, ,
  \end{eqnarray}
  where
  \begin{eqnarray}
  \Gamma_j &=& \frac{k_1^2}{k_j}\frac{\cos^2 k_j d}{2 M_j}\nonumber\\
  \Gamma_{jl} &=& \frac{k_1^2}{\sqrt{k_jk_l}} \frac{\cos k_j d\cos k_l d}{4\sqrt{M_jM_l}}\, .
  \end{eqnarray}
 The solutions of these equations are linear combinations of exponentials $e^{\alpha\Gamma t}$ with
 four possible values for $\Gamma$:
 \begin{equation}\label{rate2modes}
 \Gamma=\frac{1}{2}\left (\pm \Gamma_j\pm\sqrt{\Gamma_j^2- 4 \Gamma_{jl}^2}\right)    \, . 
 \end{equation}  
 The biggest real part of these four values determines the rate of particle creation in both modes $k_l$ and $k_j$.
 
 \subsection {Equidistant spectrum}

Another important situation is when Eq.(\ref{rescondition}) is satisfied by an infinite number of modes. This is the case, for example, for a scalar field in $1+1$ dimensions satisfying Dirichlet or Neumann boundary conditions, or,  more generally, when the spectrum is equidistant. The analysis of the solutions of Eqs.(\ref{eqsAyB}) in such cases 
is more involved. It has been described in detail for  Dirichlet boundary conditions in \cite{param1, Dalvit98}, using different approaches. 
The main results are that the coupling between an infinite number of modes makes the number of particles to grow quadratically or linearly in time, for short or 
long time-scales respectively, while the total energy grows exponentially.  This has been shown in the case in which the amplitude of the time dependence in the right hand side 
of the  modes equations is small.

\subsection {Beyond MSA: very long times}\label{beyondMSA}

The MSA improves the perturbative solutions, but it is not valid for extremely long times. For example, in the case in which there is a single resonant mode, the long 
time growth of the mode should induce an exponential behavior in all other modes. Indeed, going back to Eq.(\ref{eqmodecomp}), and assuming that the first mode is the resonant
one, we have for $n\neq 1$:
\begin{equation}
\ddot q_n+k_n^2 q_n\simeq S_{n1}(t) q_1\simeq \alpha \bar A_n e^{\lambda_1 t/2} \sin 2 k_1 t \cos k_1 t\, ,
\end{equation}
where $\bar A_n$ is a constant of order 1.  From this equation one can  show that the number of particles in all modes will grow at very long times with a rate $\lambda_1$, 
showing oscillations around the exponential with frequency $k_1$.




\section{The numerical method and results}

In order to solve numerically the equation of motion of the $n$ modes Eq.(\ref{eqmodes}), we firstly determine the eigenfrequencies of the cavity from Eq.(\ref{spectrum}) by using a single Newton-Raphson method with an stopping error of $10^{-6}$. From Eqs.(\ref{eqmodecomp})
and (\ref{ec16}), we can perform a change of variables $\dot{q}_n = U_n$, in order to obtain a new system of equations:

\begin{eqnarray}
\dot{q}_n &=& U_n, \nonumber \\
\dot{U}_n &=& -\omega_n^2(t) q_n - \sum_m S_{n m} (t) q_m , 
\end{eqnarray}
where $S_{nm}(t)$ is defined  in Eq.(\ref{ec16}). 

The system is integrated using a fourth order Runge-Kutta numerical scheme between $t=0$ and $t_{\rm max}$. The perturbation is turned on
for times $0<t<t_F$, with $t_F<t_{\rm max}$. As we know that the unperturbed solution has the form of Eq.(\ref{ec17}), 
$q_n = A_n(t)  \exp(i k_n t)/\sqrt{2 k_n}+ B_n(t) \exp(-i k_n t)/\sqrt{2k_n}$,
we can multiply both terms of the equation by $\exp(i k_n t)$ and take the mean value in $t_F <t<t_{\rm max}$. In this way, we are able to numerically evaluate $|B_n|^2$ and, also the particle number in mode $n$ as a function of time as $N_n(t) = |B_n(t)|^2/2k_n$. In our units, the 
spectral modes $k_n$ are given in units of $1/d$ ($k_n d$ is dimensionless) and consequently time is measured in units of $d$. All figures are
 referred to dimensionless quantities.

\subsection {Frequency Spectrum of tunable cavity}

Given the strong dependence of the particle creation rate with the spectrum of the static cavity, as can be seen in Eq.(\ref{onemode}) for the one resonant mode example, it is important to analyze the spectra that
result from the generalized boundary conditions in the tunable superconducting cavity (Eq.(\ref{spectrum})). Firstly, we re-define 
variables $V_0= (2 E_J)/(E_{L,{\rm cav}})$ and $\chi_0=(2 C_J)/(C_0 d)$, for which the equation reads

\begin{equation}
k_n d \tan(k_n d) + \chi_0 (k_n d)^2 = V_0 \cos f_0.
\label{spectrum2}
\end{equation}

\begin{figure}[h!]
\begin{center}
\includegraphics[width=8.5cm]{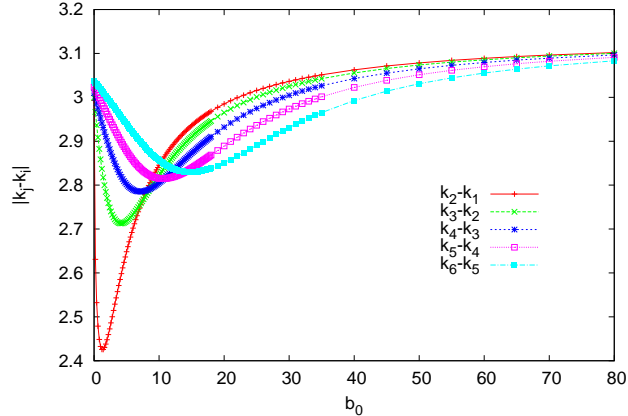}
\caption{Difference of consecutive eigenfrequencies as a function of $b_0 = V_0 \cos f_0$, for a fixed value of $\chi_0=0.05$.}
\label{fig1}
\end{center}
\end{figure}

There are three free parameters that determine the solutions of Eq.(\ref{spectrum2}): $\chi_0$, $V_0$, and $f_0$. In the first place, we shall study the difference between consecutive eigenfrequencies as a function of $b_0 = V_0 \cos f_0$  for a typical experimental value \cite{Shumeiko}, say $\chi_0 = 0.05$. 
We can see in Fig.\ref{fig1} that the bigger the value of $b_0$, the more equidistant is the spectrum. The difference between any consecutive eigenvalues of the cavity goes to a constant value of the order of $\pi$. 

\begin{figure}[h!]
\begin{center}
\includegraphics[width=14cm]{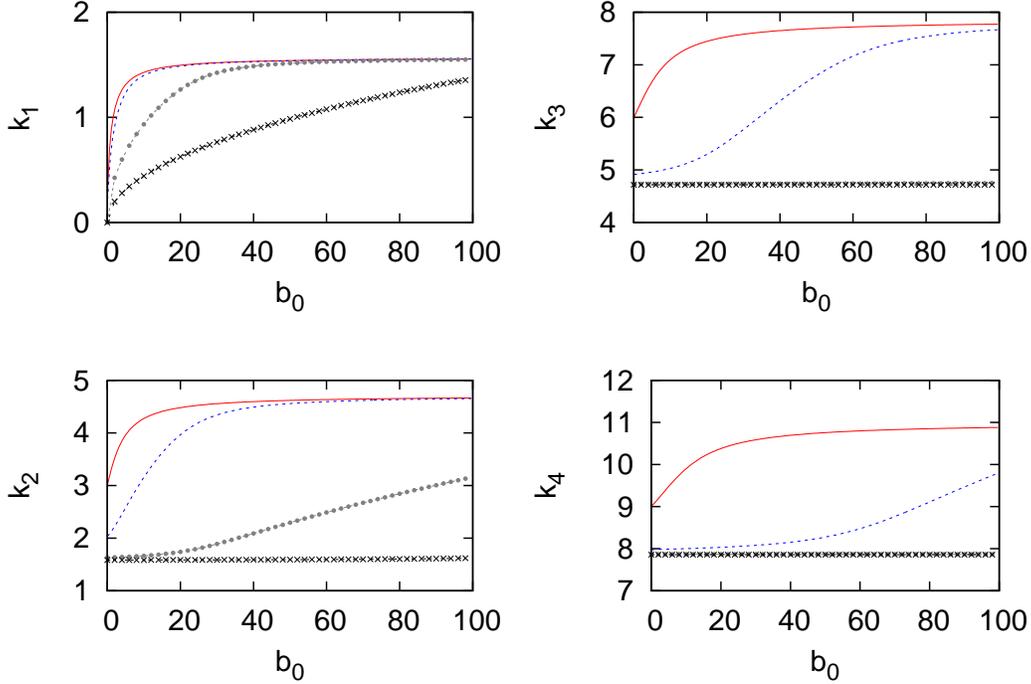}
\caption{First eigenfrequencies as a function of $b_0$ for different values of $\chi_0$.  Red-solid line is for $\chi_0=0.05$, 
blue-dashed line is for $\chi_0=1.0$, 
dotted-gray line is for $\chi_0=10.0$, 
asterisk-black line for $\chi_0=50$.}
\label{fig1bis}
\end{center}
\end{figure}

One can also study the spectrum for different values of $\chi_0$. In Fig.\ref{fig1bis} we show the first four eigenfrequencies of 
the cavity for several values of $\chi_0$. The value of $\chi_0$ determines the asymptotic value of the eigenfrequency as a function of $b_0$. 

By examining the spectrum of the cavity, we can determine different cases to be analyze carefully. Hence, we shall consider firstly, the case  
in which the eigenfrequencies are 
not spaced equidistantly, for instance the case of having one resonant mode under parametric amplification. In the end, we shall also study the case of equidistantly spanned eigenfrequencies for low or large amplitude of the driven perturbation.

\subsection {The non-equidistant spectrum and parametric amplification}

\begin{figure}[h!]
\begin{center}
\includegraphics[width=8.5cm]{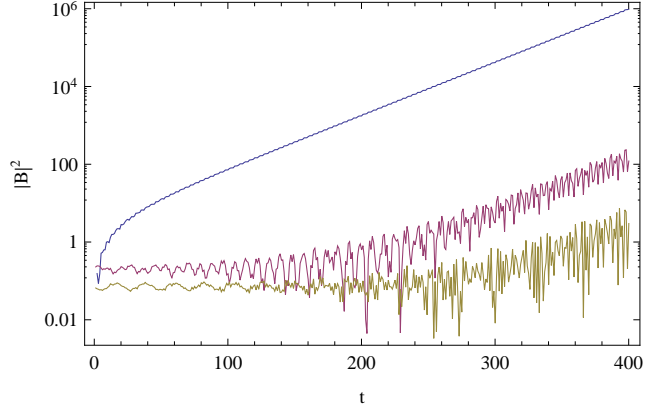}
\caption{Log-plot for $\vert B_n\vert^2$  as a function of dimensionless time for each mode of the field for a short temporal scale. 
Herein, we consider 10 modes and excite the system by $\Omega= 2 k_1$.
The three first eigenfrequencies are: $k_1= 0.8495$ (blue line), $k_2=3.2819$ (red line), $k_3= 6.1403$ (green line).}
\label{fig2}
\end{center}
\end{figure}

We shall consider values of parameters for which the spectrum is non-equidistant.  We will drive the system cavity with an external 
frequency $\Omega= 2k_1$, where $k_1$ is the first eigenfrequency of the static tunable cavity and $\Omega$ in the driven external frequency  appearing in Eq.(\ref{pert}). 
In this situation, we set $b_0 = V_0 \cos f_0=1.0$, $\chi_0=0.05$ and $\epsilon=0.001$ and we present the results for the first three modes in Fig.\ref{fig2}. It is important to note that $\Omega \neq |k_i \pm k_j |$, and therefore 
there is only a single mode under parametric resonance. As one expects, if the only resonant mode is the one tuned with the 
external frequency $\Omega = 2 k_1$, the number of created particles in this mode grows exponentially in time. The other  modes, such as   $k_2$, $k_3$, are not exponentially excited for relatively short times.  If we perform a linear fit for the log-plot of the particle number,
we obtain a value of $m_1=0.0315$ for the slope of the straight line of mode 1, in excellent agreement with the analytical prediction of Eq.(\ref{onemode}).

\begin{figure}[h!]
\begin{center}
\includegraphics[width=8.5cm]{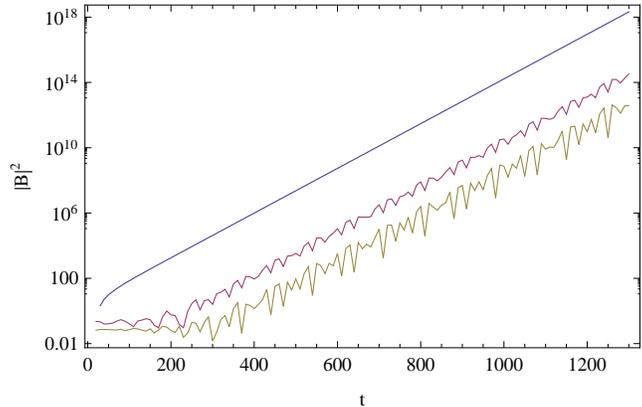}
\caption{Log-plot for $\vert B_i\vert^2$  as a function of dimensionless time for each mode of the field for a longer temporal scale. 
As in the previous figure, we consider 10 modes and the system is perturbed by $\Omega= 2 k_1$.
The three first eigenfrequencies are again: $k_1= 0.8495$ (blue line), $k_2=3.2819$ (red line), $k_3= 6.1403$ (green line).}
\label{fig3}
\end{center}
\end{figure}

\begin{figure}[h!]
\begin{center}
\includegraphics[width=8.5cm]{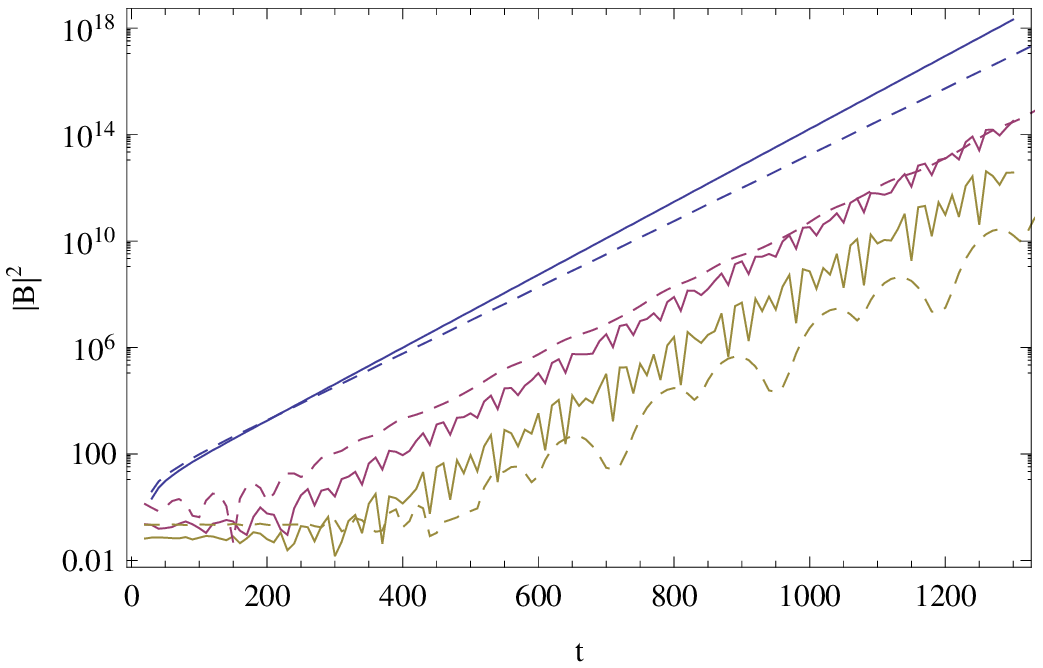}
\caption{Log-plot for  $\vert B_1\vert^2$, $\vert B_2\vert ^2$ and $\vert B_3\vert^2$. Comparison between the situations 
where $\chi_0=0.05$ (solid) and $\chi_0=1$ (dashed).}
\label{fig4}
\end{center}
\end{figure}

In Fig. \ref{fig3} we show the number of created particles as a function of dimensionless time for each mode of the field as in the previous figure, but for a longer temporal scale.  We see that at long times all modes grows exponentially with approximately the same rate. This numerical result goes beyond
MSA, and can be analytically understood  as described in Section \ref{beyondMSA}.

We can also evaluate the evolution of the system by changing the value of the parameter $\chi_0=1.0$ in Fig.\ref{fig4}. This leads to a different set of eigenfrequencies, but a similar behavior of the modes. A linear fit of the field mode 1 ($k_1=0.6799$), this time yields $m_1'=0.0289$, which means a smaller slope for bigger values of $\chi_0$. Once again, the exponential growth  of the resonant mode is well described by Eq.(\ref{onemode}), which yields an analytical value of $m_1'=0.030$.

\begin{figure}[h!]
\begin{center}
\includegraphics[width=8.5cm]{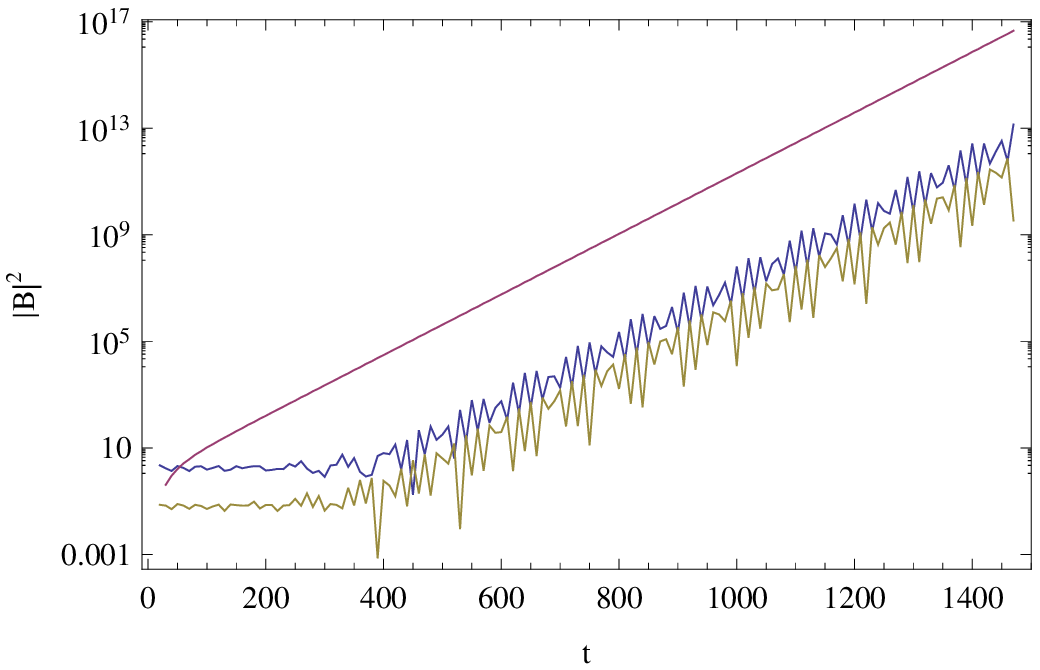}
\caption{Log-plot for $\vert B_i\vert^2$  as a function of time for each mode of the field for a longer temporal scale. Herein, we consider 10 modes and drive the system by $\Omega= 2 k_2$. The red line corresponds to $\vert B_2\vert^2$, the blue line to $\vert B_1\vert^2$ and yellow to $\vert B_3\vert^2$. Mode 2 (red line) is the more excited.}
\label{fig5}
\end{center}
\end{figure}

\begin{figure}[h!]
\begin{center}
\includegraphics[width=8.5cm]{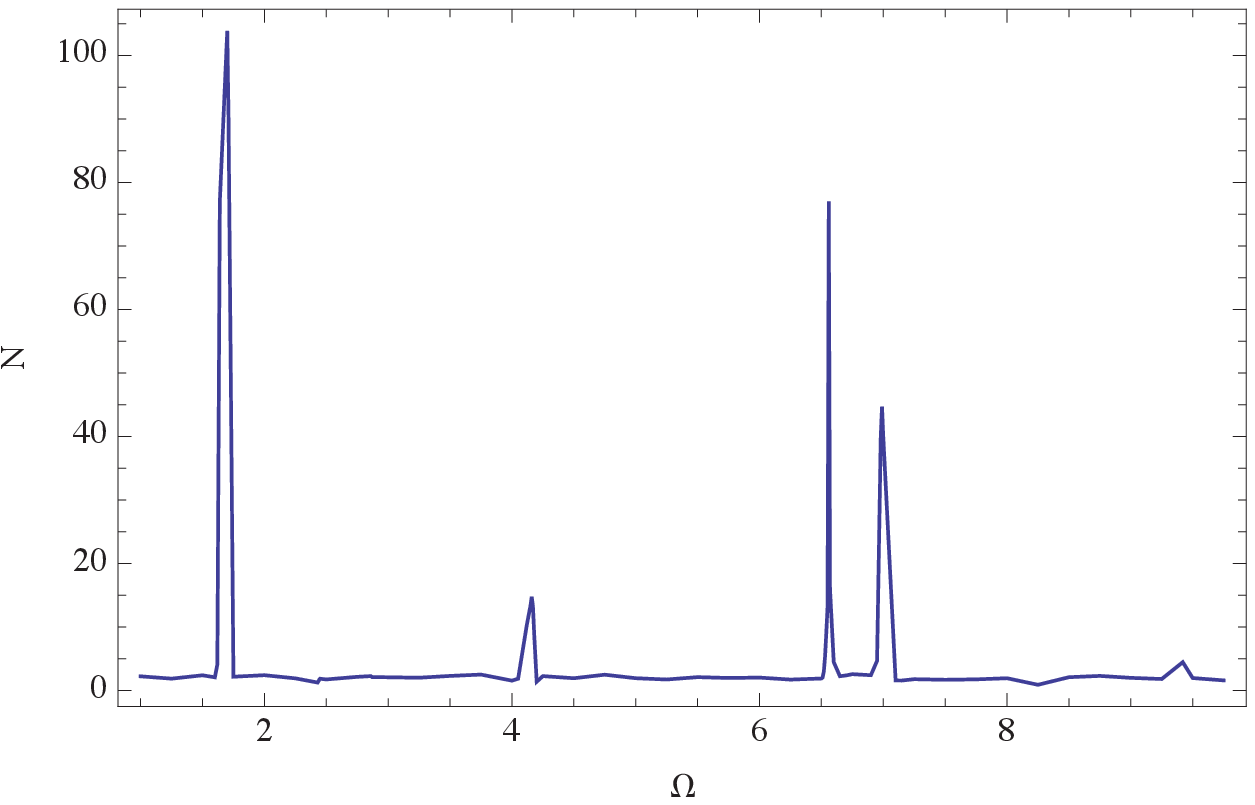}
\caption{Number of particles created as function of the external frequency $\Omega$. The  eigenfrequencies are: $k_1= 0.8495$, $k_2=3.2819$, $k_3= 6.1403$, and $k_4=9.0930$. We can see that the higher peak corresponds to $\Omega= 2 k_1$ and the following to $\Omega=2 k_2$. Less important are $\Omega= k_1+ k_2$ and $\Omega= k_1+ k_3$.}
\label{fig6}
\end{center}
\end{figure}

In Fig.\ref{fig5} we plot the number of created particles for the case in which the excitation frequency is now $\Omega = 2 k_2$. In this 
case, one can numerically evaluate the slope of the Log-plot line, obtaining: $m_2= 0.0265$ when driving  mode 2. The analytical result of Eq.(\ref{onemode}) for this  case is $m_2=0.0261$.

We performed the same simulations but for a bigger quantity of field modes. We have checked that the results are  
similar to the ones in Fig.\ref{fig2} and Fig.\ref{fig3}. For example, is we consider a cutoff of 25 field modes, the linear fit yields  fitting a slope of $m_1=0.0315$ when driving with $\Omega = 2 k_1$, similar to the one found in Fig.\ref{fig3}.

Finally, we also compute the number of particles $N$ as function of the external frequency $\Omega$ in Fig.\ref{fig6}. The  set of lower eigenfrequencies are: $k_1= 0.8495$, $k_2=3.2819$, $k_3= 6.1403$, $k_4=9.0930$. Therein, we can observe that the highest peak corresponds to an external drive of $\Omega= 2 k_1$, while the following peak refers to $\Omega=2 k_2$.  We can also note less important peaks corresponding to $\Omega= k_1+ k_2$ and $\Omega=k_1+ k_3$.

\subsection {Non-equidistant spectrum with several parametrically resonant modes}

Herein, we consider the situation in which $k_l = 3 k_j$.  If we set the parameter values as $b_0 = 4.96$, $\chi_0=0.01$, and $\epsilon=0.001$, we obtain the following eigenfrequencies: $k_1= 1.311$, $k_2 = 4.015$, $k_3= 6.862$, and $k_4=9.810$ for example. In this case, we see that 
$k_2\simeq3 k_1$, so if we excite the system with $\Omega= k_2-k_1=2 k_1$, we expect to see exponential behavior in both field modes 1 and 2. Hence,  these modes are parametrically excited,  but with a rate that takes into account the coupling between the modes.

\begin{figure}[h!]
\begin{center}
\includegraphics[width=8.5cm]{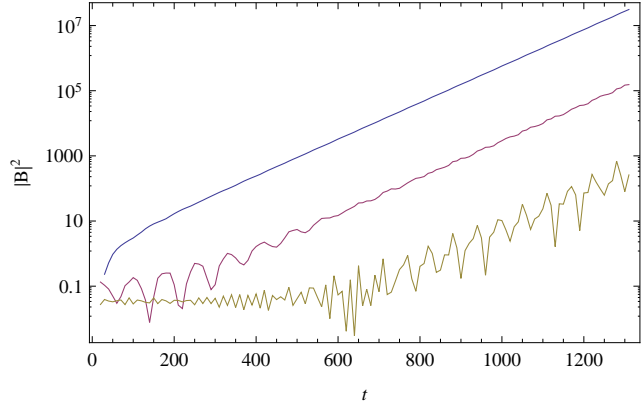}
\caption{Log-plot for $\vert B_n\vert^2$  as a function of time for each mode of the field. Herein, we consider 10 modes and excite the system by $\Omega= 2 k_1$. The blue line corresponds to the field mode 1: $\vert B_1\vert^2$, the red line to field mode 2: $\vert B_2\vert^2$ 
and the yellow one to $\vert B_3\vert^2$ for the field mode 3.}
\label{caso2m2}
\end{center}
\end{figure}

The numerical results are shown in Fig.\ref{caso2m2}. Therein, we observe the exponential grow of modes 1 and 2, as predicted by MSA. 
A linear fit yields a value of $m_1\simeq m_2 = 0.0128$ for the slope.
Likewise, an analytical prediction can be obtained by the real part of $2 \Gamma$ in  Eq.(\ref{rate2modes}), which gives $m_1=m_2=0.013$. The imaginary part in $\Gamma$ explains the oscillating behavior of the solutions, which is more pronounced for mode 2. As in the previous case, the rest of the modes start growing with the same
rate at longer times.

\subsection {Equidistantly spanned spectrum. Low amplitude perturbation}

\begin{figure}[h!]
\begin{center}
\includegraphics[width=8.5cm]{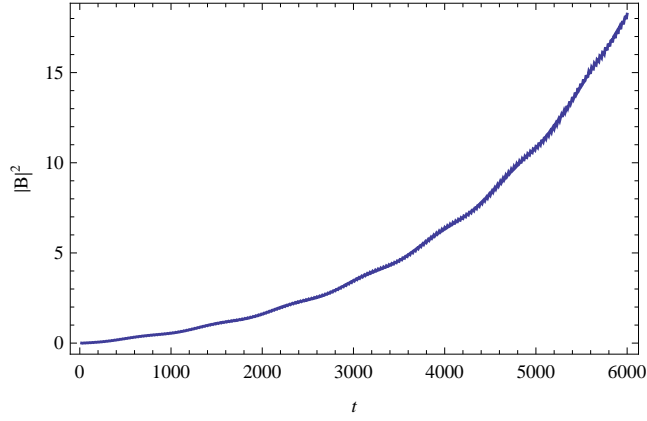}
\caption{Evolution of field mode 1 ($\vert B_1\vert^2$) for short time scale under external driving $\Omega= 2 k_1$. We can fit this behaviour with a cuadratic function.}
\label{caso3m1c}
\end{center}
\end{figure}

\begin{figure}[h!]
\begin{center}
\includegraphics[width=8.5cm]{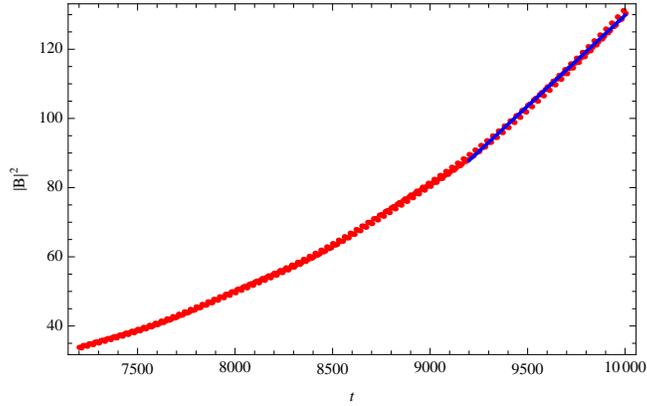}
\caption{Evolution of field mode 1 ($\vert B_1\vert^2$) for longer time scale under external driving $\Omega= 2k_1$. We can fit this behavior with a linear function for large times as can be seen in the blue line.}
\label{caso3m1L}
\end{center}
\end{figure}

In this case, we set $b_0 = 14.14$ and $\epsilon=0.005$, which corresponds to modes in the equidistant part of the cavity spectrum.  For equidistant spectra, the coupling between an infinite number of modes generates a quadratic or linear growth of the number of particles, for short or long time-scales respectively. In this case, it has already been shown that the total energy grows exponentially \cite{param1}.  With this choice,  the amplitude of the perturbation in the mode equation is given by $b_0 \epsilon \sim 0.02$, for which the MSA still applies. We see this in Figs.\ref{caso3m1c} and \ref{caso3m1L}, respectively.

\subsection {Equidistant spectrum. Large amplitude perturbation}

\begin{figure}[h!]
\begin{center}
\includegraphics[width=8.5cm]{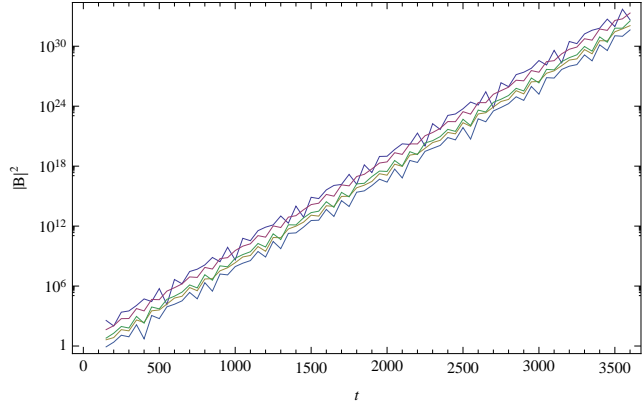}
\caption{Evolution of field modes  ($\vert B_i\vert^2$)  under external perturbation $\Omega= k_i - k_j=3.13$. We can fit this behaviour with a linear function, which indicates exponential growing of particle numbers.}
\label{caso3m1l}
\end{center}
\end{figure}

Now we look at the case of an equidistant frequency spectrum with $b_0 = 350$, $\chi_0=0.05$,  and $\epsilon=0.5$. In this case,  the amplitude of the perturbation in the mode equation is given by $b_0 \epsilon \sim 1.54$. Such a driving if beyond the MSA treatment. As expected,  we find a different behavior of modes with respect to the previous example, even in the equidistantly spanned part of the spectrum. In this case,  we report an exponentially growing number of created particles when driving with external frequency $\Omega = k_i - k_j\approx \pi$, as can be seen in Fig.\ref{caso3m1l}.

\section{Conclusions}\label{sec:conc}

In this paper we have presented a detailed numerical analysis of the particle creation for a quantum field in the presence of boundary conditions that involve a time-dependent linear combination of the field and its spatial and time derivatives. We have evaluated numerically the Bogoliubov transformation between {\it in} and {\it out}-states and found that the rate of particle production strongly depends on whether the spectrum of the unperturbed cavity is equidistant or not, and also on the amplitude of the temporal oscillations of the boundary conditions. We have provided some analytical justifications, based on MSA,  for the different regimes found numerically and emphasized the dependence of the results with the main characteristics of the spectrum.  

Firstly, we have considered a parameter set such that the spectrum of the cavity is non-equidistant. 
In this case, we have numerically solved the problem driving the system with an external frequency 
given by twice the first eigenvalue of the unperturbed cavity. As expected from MSA results, 
if the only resonant mode is the one tuned with the external frequency, the number of created particles 
in this mode grows exponentially with time. Other modes  are, for relatively short times, not 
exponentially excited.   However, at longer times, all modes grow exponentially with the same rate, a
result that goes beyond MSA. In addition, we have also considered 
a situation in which parametric resonance involved two modes. We have showed that both modes grow exponentially with a common rate that
takes into account the intermode coupling. As in the previous case, all modes are exponentially amplified at longer times.

For equidistant spectra, we have shown that the coupling between an infinite number of modes makes the number of particles  
grow quadratically in time at short timescales, and linearly in the long time limit. However, when 
the amplitude of the 
perturbations in the mode equation is large enough, the behavior of modes is qualitatively different. Hence, we have shown the existence of 
exponentially growing number of the created particles, when driving the system with an external frequency that is 
tuned at the difference between consecutive eigenvalues of the tunable cavity.  

There are several interesting issues related to the present  work which deserve further analysis. The case of a cavity ended 
by two SQUIDs may introduce interference in the particle creation rate, as in the case of two moving mirrors \cite{Dalvit2}. 
In relation to  eventual variants of recent experiments \cite{Wilson,Paraoanu}, a theoretical analysis
including nonlinearities is also due.

\section*{Acknowledgements}
This work was supported by ANPCyT, CONICET, UBA and UNCuyo. FCL acknowledges International Centre for Theoretical Physics 
and Simons Associate Program. 


\begin{thebibliography}{bib}
\bibitem{aclar} We inserted $k_1^2$ in the definition of $S_{mn}$ in order to make $\alpha$ a dimensionless constant.
\end{thebibliography}
\end{document}